\documentclass[prl,twocolumn,showpacs,amsmath,amssymb]{revtex4}

\usepackage{graphicx}
\usepackage{dcolumn}
\usepackage{bm}
\usepackage{epsfig}
\usepackage{amsmath}
\usepackage{color}
\usepackage{pifont}
\usepackage{amsmath}
\usepackage{color}

\begin{document}

\newcommand{\gam}{\gamma}
\newcommand{\kap}{\kappa}

\newcommand{\wc}{\omega_{\rm c}}
\newcommand{\wl}{\omega_{\rm l}}
\newcommand{\Da}{\Delta_{\rm {ac}}}
\newcommand{\Dac}{\Delta_{\rm {ac}}}
\newcommand{\Dc}{\Delta_{\rm c}}
\newcommand{\Ds}{\Delta_{\rm s}}

\newcommand{\TEMOO}{\ensuremath{{\textrm {TEM}}_{00} }}
\newcommand{\TEMOI}{\ensuremath{{\textrm {TEM}}_{01} }}
\newcommand{\TEMIO}{\ensuremath{{\textrm {TEM}}_{10} }}
\newcommand{\TEMIOOI}{\ensuremath{{\textrm {TEM}}_{10+01} }}
\newcommand{\FSR}{FSR}

\newcommand{\dbtpi}[1]{\ensuremath{ #1 /2\pi }}
\newcommand{\expect}[1]{\ensuremath{\left\langle #1 \right\rangle}}
\newcommand{\abs}[1]{\ensuremath{|#1|}}

\newcommand{\unit}[1]{\ensuremath{\,\mathrm{#1}}}
\newcommand{\mycite}[1]{~\cite{#1}}

\newcommand{\mytextbf}[1]{\textrm{#1}}
\newcommand{\myquotes}[1]{``#1''}

\title{Trapping and observing single atoms in the dark}
\author{T. Puppe}
\author{I. Schuster}
\author{A. Grothe}
\author{A. Kubanek}
\author{K. Murr}
\author{P.W.H. Pinkse}
\author{G. Rempe}
\email{gerhard.rempe@mpq.mpg.de} \affiliation{Max-Planck-Institut
f\"ur Quantenoptik, Hans-Kopfermann-Str.~1, D-85748 Garching,
Germany}

\date{\today}

\begin{abstract}
A single atom strongly coupled to a cavity mode is stored by
three-dimensional confinement in blue-detuned cavity modes of
different longitudinal and transverse order. The vanishing light
intensity at the trap center reduces the light shift of all atomic
energy levels. This is exploited to detect a single atom by means of
a dispersive measurement with $95 \unit{\%}$ confidence in $10
\unit{\mu s}$, limited by the photon-detection efficiency. As the
atom switches resonant cavity transmission into cavity reflection,
the atom can be detected while scattering about one photon.
\end{abstract}

\pacs{32.80.-t, 32.80.Pj, 42.50.-p}
\maketitle

A single atom coupled to a single mode of a high-finesse optical
cavity constitutes an ideal system for the investigation of
matter-light interaction at the level of individual quanta. Any
application of this system, e.g. in quantum information
science~\cite{Monroe02}, relies on the ability to precisely localize
the atom within the cavity mode at regions of strong atom-cavity
coupling. An established tool to reach this goal is an optical
dipole-force trap~\cite{Grimm00}. So far, only traps with lasers red
detuned from the atom have been demonstrated in cavity quantum
electrodynamics (QED)~\cite{Ye99, Sauer04, Maunz04, Nussmann05-2}.
In such a red-detuned dipole trap, the atom is attracted towards
intensity maxima. While this has the advantage that a single light
mode is sufficient, it has the disadvantage that the high laser
intensity perturbs the atom. As a consequence, the dynamic Stark
effect shifts the atomic energy levels and, hence, the transition
frequency. Moreover, increasing the trap depth for better
localization of the atom will increase the Stark shift. For some
atoms, like cesium, the Stark shift of a particular transition {\it
between} certain energy levels vanishes in a red dipole trap at a
magic wavelength~\cite{McKeever03}. An alternative approach is to
use a blue-detuned light field for trapping. Here the atom is
repelled from the high-intensity region and, hence, trapped close to
an intensity minimum. Such a trap has the advantage that the Stark
shift of {\it all} states can be very small so that the free-space
properties of the atom are largely retained. Superposition states,
for example, will be less affected by the trapping potential.

In this Letter we report on trapping single rubidium atoms in a
blue-detuned intracavity dipole trap. We show that the Stark shift
of the atomic transition vanishes while the atom is strongly coupled
to a cavity mode. The blue trap allows us to explore the regime of
dispersive single-atom observation. As a proof of principle, we
demonstrate that the atom is efficiently detected while scattering
only a few spontaneous photons.

\begin{figure}[b]
\begin{center}
\epsfig{file=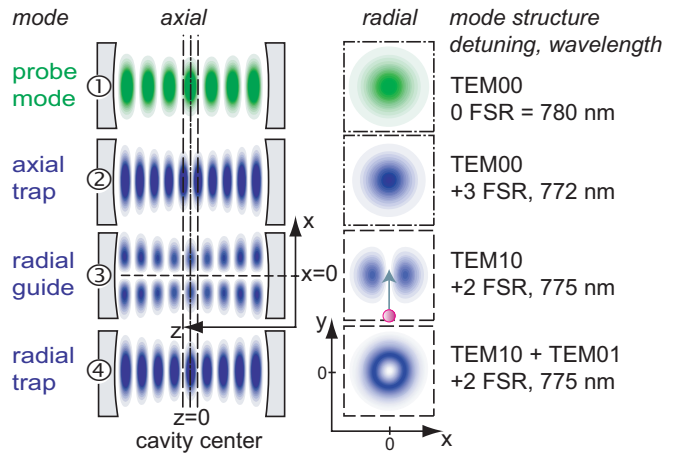,width=1\columnwidth} \caption{\small(Color
online) \mytextbf{The blue intracavity dipole trap from the
perspective of an entering atom.} The slow atom is restricted to the
field minima of the blue light fields that coincide with the
antinodes of the near-resonant probe mode \Pisymbol{pzd}{192} at the
cavity center. Persistent {\it axial} confinement is provided by a
\TEMOO\ mode \Pisymbol{pzd}{193}, \myquotes{pancakes}. Combined with
the {\it transverse} nodal line of a \TEMIO\ mode
\Pisymbol{pzd}{194}, \myquotes{funnels} are formed to guide the atom
to a strong-coupling region. Full three-dimensional confinement is
achieved by adding a \TEMOI\ mode to complete a transverse
\myquotes{doughnut} mode \Pisymbol{pzd}{195}.} \label{trap}
\end{center}
\end{figure}

The idea of the blue trap is to use far-detuned cavity modes to
shape a potential landscape which realizes three-dimensional
confinement around a dark trap center (Fig.~\ref{trap}). The
standing wave of the high-finesse cavity guarantees maximum contrast
of the interference pattern. The trap center is therefore accurately
dark. Such a blue trap has a number of advantages for experiments in
cavity QED: (1) Since the trap height does not contribute to the
atomic detuning, it can be made large for good confinement. (2) An
atom inside the trap is well isolated by the surrounding potential
barrier, outside atoms are repelled. (3) The blue trap can be loaded
by creating a dark funnel to guide a slow atom to the trap center.
As the atom is repelled from the blue light, the kinetic energy does
not increase during the capture process. Moreover, weakly coupled
atoms that are not collected by the funnel are rejected. (4) The
funnel can be closed upon detection of the strongly coupled atom in
the trap center. Because the energy gain due to guiding and
switching is kept small, the requirement to cool the atom after the
capture process is relaxed. (5) Since during the whole loading
sequence the atomic detuning is preserved, parameter regimes of
large cavity-enhanced heating~\cite{Hechenblaikner98, Murr06-2} can
be avoided.

\begin{figure}[b]
\begin{center}
\epsfig{file=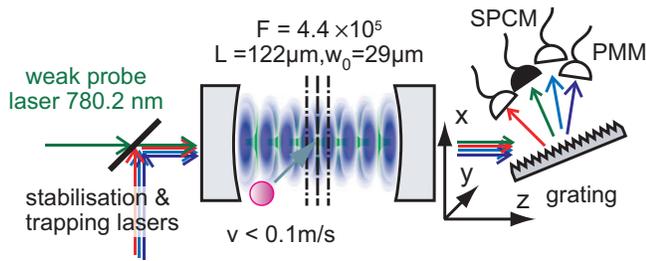,width=1\columnwidth} \caption{\small(Color
online) \mytextbf{Experimental setup.} Slow $^{85}$Rb atoms are
injected from below into a high-finesse cavity. The cavity is
excited by a weak near-resonant probe field at $780.2 \unit{nm}$ and
strong blue-detuned dipole fields. Behind the cavity, a grating
directs the probe light onto single-photon counting modules (SPCM)
whereas all dipole beams are detected by photomultiplier modules
(PMM). The cavity length $L = 0.122\unit{mm}$ is independently
stabilized by a weak laser at $785.2 \unit{nm}$. Midway between the
mirrors, the trap center coincides with an antinode of the probe
field. The mode profiles in the dash-dotted and the dashed central
plane are discussed in Fig.~\ref{trap}. The waist of the $780.2
\unit{nm}$ fundamental transverse mode is $w_0 = 29 \unit{\mu m}$,
and $F = 4.4 \times 10^5$ is the cavity finesse. The maximum
atom-cavity coupling constant and the decay rates of the atomic
polarization and the cavity field are $\dbtpi{(g_{0},\gamma,\kappa)}
= (16,3,1.4) \unit{MHz}$, respectively.} \label{system}
\end{center}
\end{figure}

The experimental setup presented in Fig.~\ref{system} is an
extension of the one described in detail
elsewhere~\cite{Muenstermann99, Maunz05}. The intracavity dipole
potential is created by a combination of standing-wave cavity modes
of different longitudinal and transverse mode order
(Fig.~\ref{trap}), all blue detuned with respect to the
near-resonant cavity QED probe field: persistent axial confinement
along the cavity axis is provided by a \TEMOO\ mode
\Pisymbol{pzd}{193} detuned by an odd number of free-spectral ranges
(FSR). The oblate antinodes (\myquotes{pancakes}) of this mode
confine the atom to the nodal planes that overlap with the antinodes
of the probe mode \Pisymbol{pzd}{192} halfway between the mirrors.
Radial confinement is provided by a doughnut mode
\Pisymbol{pzd}{195} formed by a combination of \TEMIO\ and \TEMOI\
modes detuned by an even number of FSR. To load an atom into the
trap, the radial confinement is relaxed by using the \TEMIO\ mode
only. Slow atoms from an atomic fountain are injected from below
along the $y$-direction. They are guided towards the cavity center
at $x=0$ along the nodal line of this \TEMIO\ mode
\Pisymbol{pzd}{194}. The combination of axial confinement and
transverse guiding creates funnels that direct the atom to the
antinodes of the probe mode \Pisymbol{pzd}{192}. Note that the axial
confinement need not be switched to close the trap. Since the axial
and radial characteristics of the trap are defined by independent
modes at different frequencies they can be controlled individually.

\begin{figure}[b]
\begin{center}
\epsfig{file=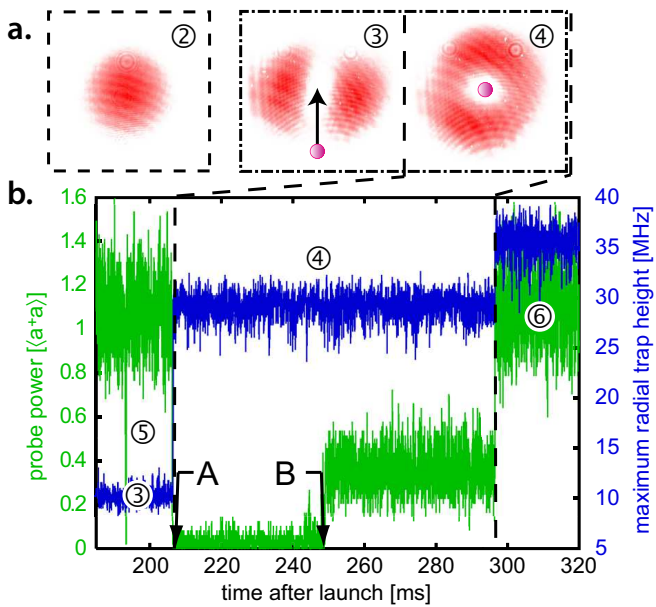,width=1\columnwidth} \caption{\small(Color
online) \mytextbf{A blue-trapping event. a). Experimental intensity
patterns} of the different modes (cp. Fig.~\ref{trap}).
\mytextbf{b). Sample trace: Transmitted probe power in units of
intracavity photon number, $\expect{a^+a}$, and the maximum radial
trap height. Upon detection of an atom (A) the probe intensity is
decreased and the trap is closed. When the atom leaves (B), the
empty cavity transmission is observed. Experimentally, the
\myquotes{doughnut} mode \Pisymbol{pzd}{195} is a controlled
superposition of two non-degenerate eigenmodes defined by our
cavity. The trace is taken for detunings of continuous axial cavity
cooling $(\Dac, \Dc) / (2 \pi) = (-35,0) \unit{MHz}$, where the
average storage time for single trapped atoms is about $30
\unit{ms}$.}} \label{trace}
\end{center}
\end{figure}

A sample trace of a trapping event is presented in Fig.~\ref{trace}.
Shown is the cavity transmission of the near-resonant probe laser at
$780.2 \unit{nm}$ and of the blue-detuned dipole laser providing the
radial confinement. The persistent {\it axial} confinement at $772
\unit{nm}$ (3 FSR detuned from the atom) \Pisymbol{pzd}{193} amounts
to a maximum potential height of $U_a = h \times 346 \unit{MHz}$,
with Planck constant $h$. The guiding field at $775 \unit{nm}$ (2
FSR detuned from the atom) \Pisymbol{pzd}{194} produces a potential
with height $U_g = 2 h \times 10.3 \unit{MHz}$ \mycite{Bluetrap1}.
The probe laser is on resonance with the bare cavity,
$\Dc=\wl-\wc=0$. Thus, the presence of an atom detunes the cavity
from resonance and causes a decrease in the transmission. Slow atoms
are guided to regions of strong coupling and cause sharp
transmission drops \Pisymbol{pzd}{196}. The trigger is armed $t=205
\unit{ms}$ after launch of the atoms from the atomic fountain to
select late atoms arriving with velocities below $0.1 \unit{m\;
s^{-1}}$. Upon detection of a strongly coupled atom in the cavity
center (A), the atom is trapped by converting the transverse guiding
mode to a confining doughnut mode \Pisymbol{pzd}{195} with a maximum
potential height of $h \times 30 \unit{MHz}$. Simultaneously, the
probe intensity is reduced. When the atom leaves the mode, the
cavity transmission increases to the bare cavity value for the
reduced observation power (B). After each trapping event, the
stabilization of all lasers and the cavity is checked
\Pisymbol{pzd}{197}.

\begin{figure}[b]
\begin{center}
\epsfig{file=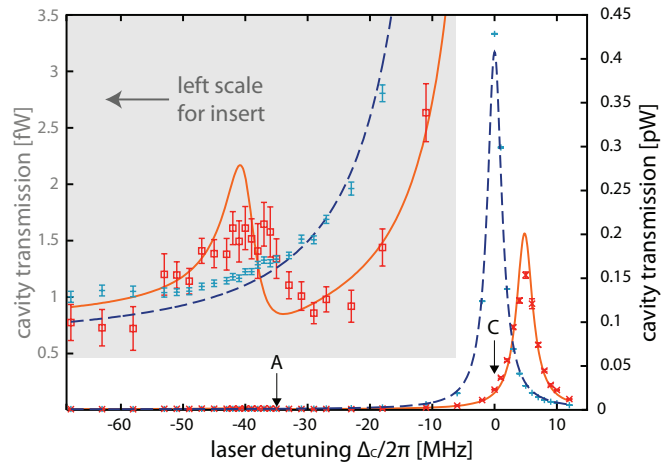,width=1\columnwidth} \caption{\small(Color
online) Normal-mode spectrum: transmission as a function of detuning
for a well-coupled system (red squares and crosses). The bare atom
(A) is detuned from the bare cavity resonance (C) by $\dbtpi{\Dac} =
-35 \unit{MHz}$. A transmission of $1 \unit{pW}$ corresponds to 1.2
intracavity photons. Intervals contribute to the spectrum if the
transmission in the neighboring cooling intervals is $<10\,$\% of
the bare cavity value $\expect{n_0}$. An analytical fit (solid line)
for a fixed coupling $g$ at low excitation results in
$g=0.83(12)\times g_0$ and a residual Stark shift of
$\dbtpi{\Delta_{\rm s}} = 0.7(1.3) \unit{MHz}$. The empty cavity
transmission (blue pluses, dashed line) is shown for reference.}
\label{nm}
\end{center}
\end{figure}

Spectroscopy of the combined atom-cavity system allows us to
determine the main characteristics of the blue trap. The
experimental protocol consists of a sequence of alternating $0.5
\unit{ms}$ long cooling and $0.1 \unit{ms}$ short probing intervals.
The probe detuning in the probe intervals is scanned with respect to
the bare cavity frequency, which is $35(1) \unit{MHz}$ blue detuned
from the atomic frequency. During the cooling intervals the probe is
on resonance with the bare cavity ($\Dc = 0$) which allows for
cavity cooling in the axial direction as well as independent
qualification of the atom-cavity coupling~\cite{Maunz04}. A probe
interval is qualified for having a strong atom-cavity coupling when
the cavity transmission in the neighboring cooling intervals is
below $10 \unit{\%}$ of the bare cavity transmission. The
expectation value of the photon number in the cavity mode is
calculated from the measured photon-detection rate and the known
detection efficiency including propagation losses from the cavity to
the detectors. Experimental results are displayed in Fig.~\ref{nm}.
We purposely chose a large atom-cavity detuning, $\Dc
> 2g_0$, with $\dbtpi{g_{0}} = 16 \unit{MHz}$ the maximum
atom-cavity coupling constant, to show that the blue trap preserves
this detuning and allows to enter the regime of dispersive
detection, as discussed below. Analytical results for an atom with
fixed coupling (solid curve) fit the data (points) well. Comparison
between theory and experiment gives an atom-cavity coupling constant
of $83(12) \unit{\%}$ of $g_{0}$, much larger than the atomic and
cavity decay rates. This proves that a strongly-coupled atom-cavity
system has been prepared.

In Fig.~\ref{nm}, the measured Lorentzian transmission peak of the
bare cavity at $\Dc = 0 \unit{MHz}$ is shown for reference. The bare
atom is at $\dbtpi{\Dac} = -35(1) \unit{MHz}$, with the uncertainty
due to some residual magnetic field. The bare detunings are the same
as those of the normal-mode experiment performed previously in a red
dipole trap~\cite{Maunz05}. The obvious difference in the
normal-mode spectra lies in the fact that in a red trap the atomic
resonance is shifted by approximately twice the ground-state trap
depth, effectively bringing the atomic transition frequency close to
resonance with the cavity at the trap center (see Fig.~2
in~\cite{Maunz05}). In contrast, the normal-mode spectrum of an atom
stored in the blue-detuned trap does not show any shift of the
atomic frequency, as expected for an atom trapped at the node of the
blue field. Since the large atom-cavity detuning is preserved, the
character of the normal modes emerging from the bare states remain
largely \myquotes{atom-like} and \myquotes{cavity-like}. The strong
asymmetry of the peak heights arises from the fact that the system
is excited via the cavity and observed in transmission. Therefore,
the atom-like resonance is much weaker than the cavity-like
resonance. The plots on shaded background are a blow-up of the
spectrum ($\sim 130 \times$) to present the atom-like peak located
at approximately $\dbtpi{\Dc}=-40 \unit{MHz}$. This peak is slightly
broadened by the spatial distribution of the atoms in the mode. A
residual Stark shift $\dbtpi{\Delta_{\rm s}} = 0.7(1.3) \unit{MHz}$
can be derived from the analytical fit~\cite{Bluetrap2}. This shift
is much smaller than the axial and radial trap heights, $U_a = h
\times 265(6) \unit{MHz}$ and $U_r = h \times 30(1) \unit{MHz}$,
respectively. The Stark shift due to the red-detuned stabilization
laser at $785.2 \unit{nm}$ is $\dbtpi{\Delta_{\rm {stab}}} = 2.2(1)
\unit{MHz}$. The shift of the atomic transition frequency due to the
blue trap is therefore smaller than the atomic linewidth.

The preservation of the large atom-cavity detuning facilitates
dispersive measurements~\cite{Quadt95,Hood98,Muenstermann99} while
the blue trap provides confinement. This is exemplified by the
detection of an atom in the cavity via the induced shift of the
cavity-like normal mode. Such a measurement scheme keeps the atomic
excitation low. To estimate the average number of spontaneously
scattered photons during a certain observation time interval, we
consider probing the system on resonance with the bare cavity. In
the presence of a strongly coupled atom, the cavity transmission of
the probe is reduced by a factor of $\sim 20$. The transmission is a
direct measure of the excitation of the mode corresponding to
$\expect{a^+ a}$ photons. In the limit of weak excitation the
excitation probability of the atom is proportional to the photon
number times the atomic Lorentzian: $\expect{\sigma^+\sigma^-} =
\expect{a^+a} g^2\big/(\Dac^2+\gam^2)$ where $a^+ (a)$ is the
creation (annihilation) operator for cavity photons, and $\sigma^+
(\sigma^-)$ is the atomic raising (lowering) operator. The atomic
excitation $\expect{\sigma^+\sigma^-}$ is therefore given by the
cavity excitation $\expect{a^+ a} = 0.022$ (cp. (C) in
Fig.~\ref{nm}) times a constant which depends on the effective
coupling, $g$, the atom-cavity detuning, $\Dac$, and the atomic
linewidth, $\gamma$. The effective coupling was deduced from the
experimental data in Fig.~\ref{nm}, and $\gamma$ and $\Dac$ are well
known. The deduced average atomic excitation of
$\expect{\sigma^+\sigma^-} = 3.1 \times 10^{-3}$ leads to a
scattering rate into free space given by $2 \gamma
\expect{\sigma^+\sigma^-} \approx 117\unit{kHz}$. Thus, during a
time interval of $10\unit{\mu s}$ the atom scatters $1.2(3)$
photons. This includes an overall experimental detection efficiency
of $5 \%$ for photons lost from the cavity mode. A detailed analysis
which assumes a $50 \%$ a priori probability of the presence of the
atom in the cavity, i.e. maximum possible ignorance, and which takes
into account the Poissonian statistics of the detected photons and
all experimental imperfections, results in a $95 \%$ correct
decision concerning the presence of the atom in this $10 \unit{\mu
s}$ long time interval. The required observation time interval
scales inversely with the photon-detection efficiency which can be
improved considerably.

While trapped, the atom heats up due to spontaneous emission and
dipole-force fluctuations~\cite{Hechenblaikner98,Murr06-2}. The
latter heating process is largely compensated by cavity
cooling~\cite{Maunz04}. For the parameters of Fig. \ref{trace}, the
average storage time is about $30 \unit{ms}$. For a different trap
height we also measured the storage time without the probe light to
be about 20 ms. Moreover, we see the same dependence on the probe
power as reported in Ref.~\cite{Maunz04}, namely a trapping time
decreasing inversely with increasing probe power. The observed
trapping times are comparable to those found in a red
trap~\cite{Maunz05} mainly because radial heating is not compensated
for.
An increase of the storage time by several orders of magnitude can
be achieved  for cavity cooling in three dimensions, which requires
illuminating the system from the side~\cite{Nussmann05-3}. In this
scheme all probe lasers would be red detuned from both normal modes,
such that light scattering increases the photon energy and, hence,
cools the atom. Since in cavity cooling photons should predominantly
be emitted via the cavity mode, this requires the lower dressed
state to be \myquotes{cavity-like}. An advantage of this cooling
scheme would be that it is efficient for a strongly coupled atom at
the trap center. Note that for a cooling laser resonant with the
bare cavity ($\Dc=0$), as is the case in the experiment underlying
Fig.\ref{nm}, cooling is achieved only for an atom close to a node.
As a first step in this direction, we have successfully captured and
stored single atoms in the blue trap for the appropriate parameters
for 3D cavity cooling.

In conclusion, we have realized a blue intracavity dipole trap which
now allows measurements in cavity QED while largely preserving the
free-space level structure of the confined atom. %
Well controlled detunings and coupling are important for the
investigation of genuine quantum effects, where a reliable coherent
evolution is essential. For applications in quantum information
science, the absence of differential energy shifts reduces dephasing
of superposition states. Further perspectives for the blue trap
include the possibility to control the atomic motion by means of
feedback~\cite{Fischer02}. Such experiments would benefit from the
possibility to independently address the radial and axial
confinement. Moreover, in experiments where optical cavities are
investigated as single-atom
detectors~\cite{Horak03,Trupke05,Oettl05,Steinmetz06,Teper06} the
blue \myquotes{funnels} demonstrated in this Letter could
efficiently guide atoms to regions of large atom-cavity coupling,
thereby enhancing the detection efficiency.

\begin{acknowledgments}
\end{acknowledgments}


\begin{thebibliography}{22}
\bibitem[{Blu({\natexlab{a}})}]{Bluetrap1}
\bibinfo{note}{The maximum radial trap height in Fig. \ref{trace} is given for
  the doughnut mode. Since it scales with the ratio of the maximum intensity to
  the integrated intensity, the trap height for the ${\textrm {TEM}}_{01}$
  guiding mode is a factor $2$ larger.}
\bibitem[{Blu({\natexlab{b}})}]{Bluetrap2}
\bibinfo{note}{A conservative upper bound for the Stark shift of
  $\dbtpi{\Delta_{\rm s}} < 4 \unit{MHz}$ can be extracted from fitting only
  the slopes of the atom-like peak.}
\end{thebibliography}
\end{document}